\documentclass[12pt]{iopart}

\usepackage{graphicx}
\usepackage{cite}
\usepackage{zref-xr,zref-user}
\usepackage{nameref,hyperref}
\zexternaldocument*{Supplementary_Information_PhysicalBiology}
\usepackage[inline]{enumitem}
\usepackage{tikz}
\usetikzlibrary{shapes}
\begin{document}

\newcommand{\markerone}{\raisebox{0.5pt}
{\tikz{\node[draw,blue,scale=0.4,regular polygon, regular polygon sides=4,fill=blue](){};}}}
\newcommand{\markerthree}{\raisebox{0.5pt}
{\tikz{\node[draw,magenta,scale=0.3,regular polygon, regular polygon sides=3,fill=magenta,rotate=0](){};}}}
\newcommand{\markerfour}{\raisebox{0.5pt}
{\tikz{\node[draw,scale=0.4,regular polygon, regular polygon sides=4,fill=none](){};}}}
\newcommand{\markerfive}{\raisebox{0.8pt}{\tikz{\node[draw,red,scale=0.4,circle,fill=red!20!red](){};}}}
\newcommand{\markersix}{\raisebox{0.6pt}{\tikz{\node[draw,scale=0.4,circle,fill=red](){};}}}
\newcommand{\markerseven}{\raisebox{0.5pt}
{\tikz{\node[draw,blue,scale=0.3,regular polygon, regular polygon sides=3,fill=blue,rotate=0](){};}}}
\newcommand{\markereight}{\raisebox{0.5pt}
{\tikz{\node[draw,black,scale=0.4,regular polygon, regular polygon sides=4,fill=black](){};}}}
\newcommand{\markerten}{\raisebox{0.5pt}
{\tikz{\node[draw,blue,scale=0.3,regular polygon, regular polygon sides=3,fill=none,rotate=0](){};}}}
\newcommand{\markereleven}{\raisebox{0.5pt}
{\tikz{\node[draw,black,scale=0.4,regular polygon, regular polygon sides=4,fill=none](){};}}}
\newcommand{\markertwelve}{\raisebox{0.8pt}{\tikz{\node[draw,red,scale=0.4,circle,fill=none](){};}}}
\newcommand{\markernine}{\raisebox{0.8pt}{\tikz{\node[draw,black,scale=0.4,circle,fill=black](){};}}}

%\title[Author guidelines for IOP Publishing journals in  \LaTeXe]
\title[]
{Aggregation dynamics of active cells on non-adhesive substrate}

\author{Debangana Mukhopadhyay\textsuperscript{1},
Rumi De\textsuperscript{1*},}

\address{Department of Physical Sciences, Indian Institute of Science Education and Research Kolkata, Mohanpur 741246, India.}
\ead{rumi.de@iiserkol.ac.in}
\vspace{10pt}
\begin{indented}
\item[]December 2018
\end{indented}

\begin{abstract}
Cellular self-assembly and organization are fundamental steps for the development of biological tissues. 
In this paper, within the framework of a cellular automata model, we address how an ordered tissue pattern spontaneously emerges from a randomly migrating single cell population without the influence of any external cues.
This model is based on  the active motility of cells and their ability to reorganize due to cell-cell cohesivity as observed in experiments.  Our model successfully emulates  the formation of  nascent clusters and also predicts the temporal evolution of aggregates that leads to the compact tissue structures.
Moreover, the simulations also capture several dynamical properties of growing aggregates, such as, the rate of cell aggregation and non-monotonic growth of the aggregate area which show a good agreement  with the existing experimental observations.  We further investigate the time evolution of the cohesive strength, and the compactness of aggregates, and also study the ruggedness of the growing structures by evaluating the fractal dimension to get insights into the complexity of tumorous tissue growth which were hitherto unexplored. 
\end{abstract}

%\maketitle

\section{Introduction}
Formation and development of tissues through self-assembly and organization of living cells are intriguing and complex phenomena \cite{Sasai2013, Whitesides2002}. It is still not well understood how ordered tissue structures spontaneously develop from orchestrated response of interacting multi-cellular components. 
Understanding the process of tissue organization, thus, would immensely benefit diverse areas of developmental biology, wound healing, cancer therapy, tissue engineering and even organ printing to name a few \cite{Sasai2013, Whitesides2002, Jakab2010,   Rodriguez2012, Mironov2009, Neagu2005, Douezan2012, Rumi_safran_2010, Rumi_FA_2018}.

In nature, numerous examples of self-assembled aggregation processes can be found both in living as well as non-living systems \cite{Jacob1990}, such as, 
formation of snowflakes \cite{Langer1980}, cloud formation \cite{Blando1990}, coagulation of colloids\cite{Weber1991},  aggregation of proteins\cite{Bence2001}, swarming of bacteria \cite{Jacob1994, Matsushita1998,Tsimring1995}, flocking of birds\cite{Vicsek2012} etc.
Several experimental, theoretical, and computer simulation studies have been carried out which reveal a great deal about the structural and dynamical aspects of self-assembly and aggregation processes in passive systems\cite{Meakin1998, Barabasi1995}.  
For an example, studies on diffusion-limited aggregations provide a deep understanding into the process of snowflakes like branched dendritic patterns formation  \cite{Witten1983, Meakin1983, Meakin1984, Vicsek1984, Kolb1983}.
Moreover, it has been found that  variety of inter-particle interactions, reaction mechanisms, coalescence rates, and other factors 
play a crucial role in determining the dynamics of aggregates \cite{Meakin1998, Barabasi1995}.

There have also been many efforts to understand the assembly and organization processes of living tissues, such as, how tissues spread \cite{Ryan2001, Douezan2011, Guo2006}, how sorting takes place in a multicellular system \cite{Jakab2004, Flenner2012, Graner1992,Beatrici2011}, or how tumors develop and grow  \cite{Rodriguez2012,Tracqui2009}.
An insight into the complex tissue organizations could be obtained  
based on differential adhesion hypothesis (DAH) proposed by Steinberg {\it et. al.} \cite{Steinberg1963, Steinberg2007, Foty2005}. According to DAH, motile and cohesive cells spontaneously tend to reorganize  to maximize cell-cell cohesive binding strength and to minimize the interfacial energy of aggregates. DAH, thus,  provides an important route to the formation of 
ordered tissue patterns from collective interactions of multi-cellular components. 
There are also quite a few theoretical and computational studies to understand the complex behaviours of tissues, for examples, Graner and Glazier have developed theoretical model to investigate the phenomenon of cell sorting \cite{Graner1992}; Sun and Wang's group have simulated fusion of multicellular aggregates \cite{Sun2013}; Flenner {\it et. al} have modelled temporal shape evolution of multicellular systems \cite{Flenner2012}.
Moreover, reaction-diffusion mechanisms have also been shown to influence the formation of many tissue patterns, such as rapidly growing embryo or stem cell aggregate \cite{Sasai2013}, aggregation of amoebae motion \cite{Vasiev1994}.

Interestingly, in recent experiments, performed by Douezan and Brochard-Wyart, it has been observed  that randomly migrating cells deposited on non-adhesive surface spontaneously 
form closely packed compact aggregates \cite{Douezan2012}. There are other experiments which also exhibit the tendency of cellular aggregates to spontaneously form compact tissue structures acquiring minimum area \cite{Guo2006, Koch1990, Hiram1983, Vicsek1985, Huang2004}. Examples include,  rounding up of Hydra aggregates  into circular shapes in $2$D \cite{Rieu2002} or chick embryonic cellular aggregates evolving into spherical shapes in $3$D \cite{Mombach2005}; in all these cases, cell-cell adhesion bonds have been found to play a significant role in shaping up the structure.
A key step to get insights into these processes is to understand what drives cell-cell attraction to form compact multi-cellular aggregates. 
%remains to be explored 
Several hypotheses have been proposed, in particular, whether cell-cell attraction is based on chemical signalling \cite{Weijer2009} or whether cellular communications are mediated by extracellular matrix - where deformations created in
the substrate by one cell can be detected by an adjacent cell \cite{Oers2014}, or the attraction is mediated due to haptotaxis, i.e., directional cell motility up a gradient in the substrate \cite{Weber2013}.
Motivated by the experimental findings, in this paper, we address how an ordered tissue pattern spontaneously emerges from a seemingly disordered single cell population without the influence of any external physical or chemical forces.
We show that a suitably constructed cellular automata model 
based on the active motility and local reorganization of cells can successfully capture the formation of  nascent clusters and predict the temporal evolution of aggregates that leads to the compact tissue structures as observed in experiments.
The crux of our model is that the sole consideration of the cellular tendency of forming bonds with  neighbouring cells to maximize cohesive strength, is sufficient for the emergence of compact tissue pattern. Our study thus shows that the presence of an external cue or a mediator or a gradient  is not critical to these types of aggregations which was hitherto unexplored.
Our theory also reveals the existence of two distinct time scales in such cellular aggregation processes - one is fast time scale associated with the diffusion of cells and another much slower time scale associated with the tissue compaction process that involves breaking of cell-cell cohesive bonds and making of new bonds.
Interestingly, we find that the difference in the tendency of cell types to self-organize to increase the binding strength plays a crucial role in determining the time scale and the structure of the tissue pattern.

Apart from emulating the structural evolution, 
our model also captures several dynamical properties of cellular aggregates \cite{Douezan2012,Rodriguez2012,Guo2006}, such as, the rate of aggregation, {\it i.e.}, how the number of cell clusters evolve with time as the aggregation takes place. 
Moreover, as observed in experiments, our model also predicts
the non-monotonic growth of the surface area of the growing aggregates on non-adhesive substrates. We further investigate, how the cellular cohesive binding strength and the overall compactness of the growing aggregate evolve as time progresses which remains so far unexplored. Besides, we also study how the ruggedness of the growing aggregate structure changes towards a smooth compact structure by evaluating the fractal dimension of the aggregate which can be useful in comparing normal versus deceased tissue growth as revealed by the recent experiments \cite{Baish2000, Campbell2016}.

\section{Cellular automata modelling}
\begin{figure}[h]
\centering
\begin{center}
\includegraphics[scale=0.6]{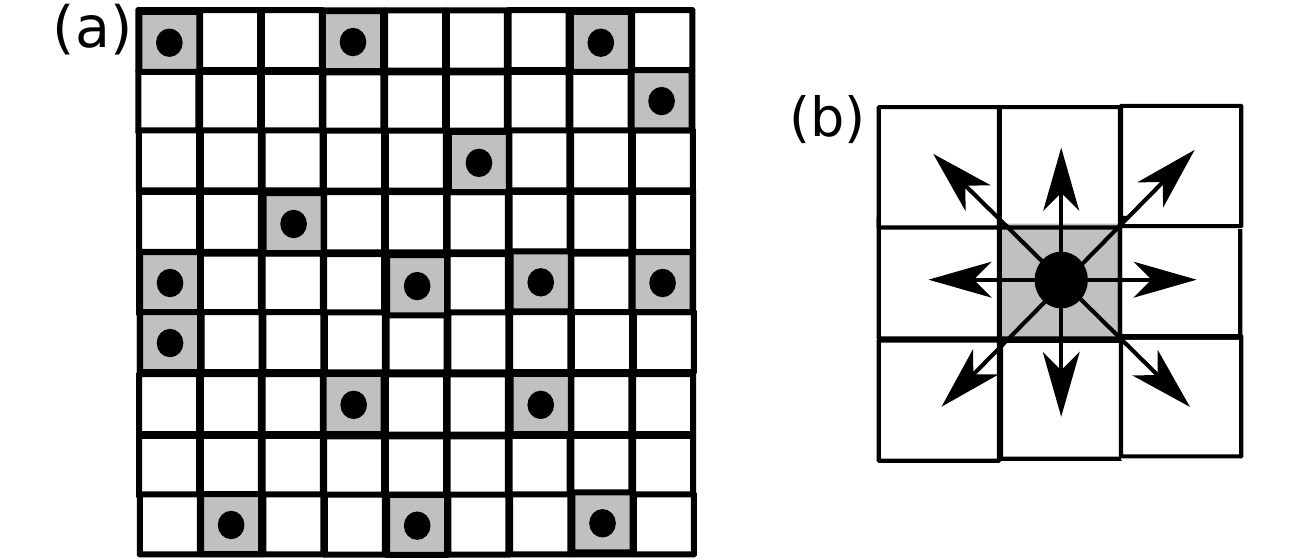}
\end{center}
\caption{(a)  Illustration of randomly deposited cells on a lattice surface. Shaded areas represent lattice sites occupied by cells and each dot represents the center of mass of the cell.(b) Schematic of eight neighbouring sites of a cell.}
\label{fig_1}
\end{figure}

In this section,  we now discuss the formulation of our cellular automata model in details.  First, we start with $N_0$ cells  randomly deposited  on  a $(L \times L)$ square lattice as illustrated in Fig. \ref{fig_1}(a). Each lattice site is, thus, either occupied by a cell or remains empty.
 Each cell is considered to have eight nearest neighbouring sites as depicted in Fig. \ref{fig_1}(b) to form a close packed structure in a square lattice. In our model, cells can diffuse to any of its empty  neighbouring sites. On the way, if it finds another immediate neighbouring cell then sticks to it due to cell-cell binding affinity and  starts moving together as a whole cluster. Here, the sticking probability is considered to be $1$.
Thus, at any instant, there are $N(t)$ number of cell clusters.
It is to be noted that, in our simulation, we refer a single cell also as a cluster. Thus, a cluster can consist of a single cell or multiple cells.  In our model,  at each simulation step, $N(t)$ clusters 
are picked up at random one by one and then the cluster is given the opportunity to (i) either diffuse with a motility probability, $P_m(n)$, (ii) or locally reorganize with a rolling probability, $P_r(n)$,  within its own cluster so that cell-cell cohesive binding strength increases, (iii) otherwise, just stay put.

{\bf Motility criteria:} 
The motility probability of a cluster is considered to be dependent on its size as $P_m(n)= \frac{\mu}{n}$; where  $n$ is the number  of cells in the cluster and $\mu$ is the proportionality constant. In our simulations, $\mu$ is taken as $1$. 
 As time progresses, diffusing clusters collide and merge to form a bigger cluster.
Thus, the  motility of the cluster decreases  with accumulation of more number of cells. 
The dependence of the probability, $P_m(n)$  on $1/n$ could  simply be attributed to the decreasing tendency of diffusion of the cluster as it gathers more mass, since the diffusion coefficient, $D \propto \frac{1}{m}$; where $m$ is the mass to the cluster and hence, proportional to the number of cells, $n$, of the cluster \cite{Phillips2013}.
Moreover,  as found in experiments, the cluster motility gradually slows down with increase in size and eventually stops moving after it reaches a critical size (say, $n_c$ number of cells).

{\bf Local reorganization:} 
Now, we discuss  in detail the  local  reorganization of cells within its own cluster via rolling process.  As cells in a cluster are bound together by strong cell-cell adhesion, local reorganization process requires breaking of existing cell-cell cohesive bonds and again making of new bonds. Thus, this process is observed to be much slower process compared to diffusion. 
Therefore, in our model, we consider an additional probability factor, $P_r(n)=\beta n$ to incorporate the slower compactification process, where the rolling  parameter $\beta$  denotes the tendency of the cell type to locally reorganize due to the binding affinity.
The cell rearranges its position relative to the neighbouring cells so that it is surrounded by the maximal possible neighbours  and hence, increases the cell-cell binding strength. As the cluster size gets bigger, probability of rolling increases with increase in number of cells, $n$, in the cluster.

In our model, the cell-cell binding interaction energy is described as,
\begin{equation}
E=-\epsilon_1 n_1 -\epsilon_2 n_2 -\epsilon_3 n_3;
\label{Eqn-interactionenergy}
\end{equation} 
where $n_1, n_2$, and $n_3$ are the number of first, second and third nearest neighbouring cells and $\epsilon_1$, $\epsilon_2$, and $\epsilon_3$  are the weight factors given to its neighbouring binding sites respectively. 
 Here, we assume that the interaction strength between a cell and its first nearest neighbours is maximum as they are adjacent to each other and the interaction strength gradually decreases from second to third neighbours.
It is also observed in experiments that cell-cell interaction can extend beyond first nearest neighbours, since apart from the physical contact, cell-cell attraction could be driven by chemical signalling i.e. cells secrete  chemoattractants into the medium.

Now, at each simulation step, for a randomly chosen cluster, if the motility criteria, $P_m(n)$, are not satisfied, then we check  for the rolling probability criterion, $P_r(n)$. Once it is satisfied then we check whether the local reorganization can take place.   
We, first, calculate the binding interaction energy, $E_0$, of the cell in its current location following Eq. (\ref{Eqn-interactionenergy}). Next, we  choose randomly an empty 1st nearest neighbouring site for its possible relocation and calculate its binding energy, $E_{n}$, at the new position.
If the corresponding binding energy, $E_{n}$, associated with the new location is lowered compare to its current configuration, $E_0$, so that it leads to an over all increase of cellular binding strength due to more cell-cell interactions, then the move is readily accepted.  
However, if the energy, $E_{n}$, becomes higher, then a random number, $r$, is generated from the uniform distribution over the interval $[0, 1]$, and the new  configuration is accepted with a probability such that $r < \exp \big( - \Delta E\big)$, where $\Delta E=E_{n}-E_{o}$, is the corresponding change in cellular interaction energy. 
In simulations, we have considered, $\epsilon_1=3$, $\epsilon_2=2$, and $\epsilon_3=1$.

\subsection{Numerical algorithm}
Details of  numerical steps of our cellular automata model  are summarized as follows:\\
%\begin{itemize}
\begin{enumerate}[label=(\roman*)]
\item $N_0$ cells  are, initially, deposited at random on a $L \times L$ two dimensional square lattice as shown in Fig. \ref{fig_1}(a).
(As single cell is also referred as a single cell cluster; thus, initially, the number of clusters, $N(t)=N_0$.)

\item At each simulation step,  $N(t)$ clusters are randomly picked up one by one (once a cell of a cluster is randomly chosen, no other cells of that particular cluster will be chosen at that step).
 Then, we record the number of cells, $n$, of the  chosen cluster. Next, we check whether the motility  or the local reorganization or the stay put criterion of the cluster is satisfied.

\item Motility criteria:  a random number $R_1$ is generated from the uniform distribution over the interval $[0,1]$. 
If  $R_1 \le P_m(n)=\frac{1}{n}$ and the size of the cluster is less than a critical number $n_c$, then the whole cluster moves in any of the empty first nearest neighbouring sites.

\item Criteria of local rearrangement :
 if the motility criteria are not satisfied then we check for the rolling criteria.  We call another random number  $R_2$ and if 
 $R_2 \le P_r(n)=\beta n$, then the chosen cell of that cluster given an opportunity to locally rearrange within its own cluster to be surrounded by more number of neighbours.

\item Now, for reorganisation process to occur, an empty site in its first nearest neighbouring region is randomly chosen.
Then, we calculate the binding energy, $E_n$, in its possible new location and
the energy, $E_0$, in its current location.
If $E_{n} \le E_o$, the new location is readily accepted. Otherwise, if $E_{n} > E_o$, then we call a random number $(r)$ from a uniform distribution $[0,1]$, if $r < \exp(-\Delta E)$ where $\Delta E = E_{n}-E_o$, then the move is accepted except if the move disintegrates the cluster then that move is not allowed
(as illustrated in Figure \zref{enegy_cal_notallowed}(a) \& (b) in supplementary information).

\item If both the motility and the rolling criteria are not satisfied, then the cluster just stays put.
\end{enumerate}

\section{Results and discussions}

We now investigate the dynamics of initiation, growth, and evolution of cellular aggregates on non interacting surface following the cellular automata model described in the previous section. 
In our simulations, we start from a population of randomly distributed single cells  spread over a $2$D surface. While diffusing on the surface, cells collide, stick to each other, and thus form small nascent clusters of two or three cells clusters.  
These clusters then grow with accumulation of more colliding cells or clusters and give rise to large aggregates. Our simulations capture many dynamical properties of growing aggregates, those we discuss in detail in the following subsections. 
We have performed simulations for a wide range of parameter values and  results
presented here after averaging over many such simulations.

\begin{figure}[!t]
\centering
\begin{minipage}[b]{0.3\textwidth}
\label{fig:cc}
%\caption{Using triangular lattice, \\A= 0.3994, D=1.601}
\begin{center}
\includegraphics[scale=0.18]{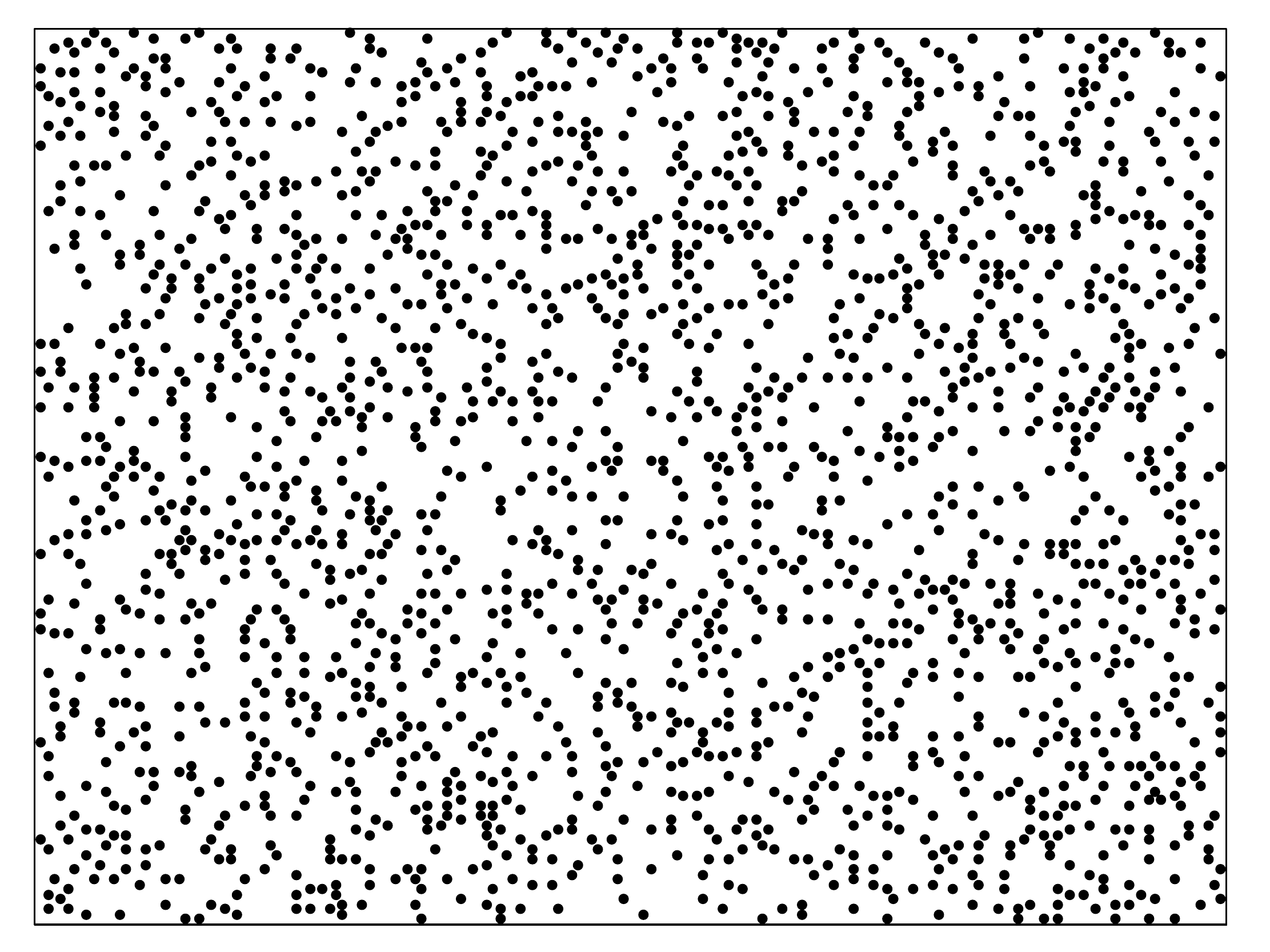}
\center {(a) T=0}
\end{center}
\end{minipage}
\begin{minipage}[b]{0.3\textwidth}
\label{fig:cc}
%\caption{Using square lattice, \\A= 0.39613, D=1.604}
\begin{center}
\includegraphics[scale=0.18]{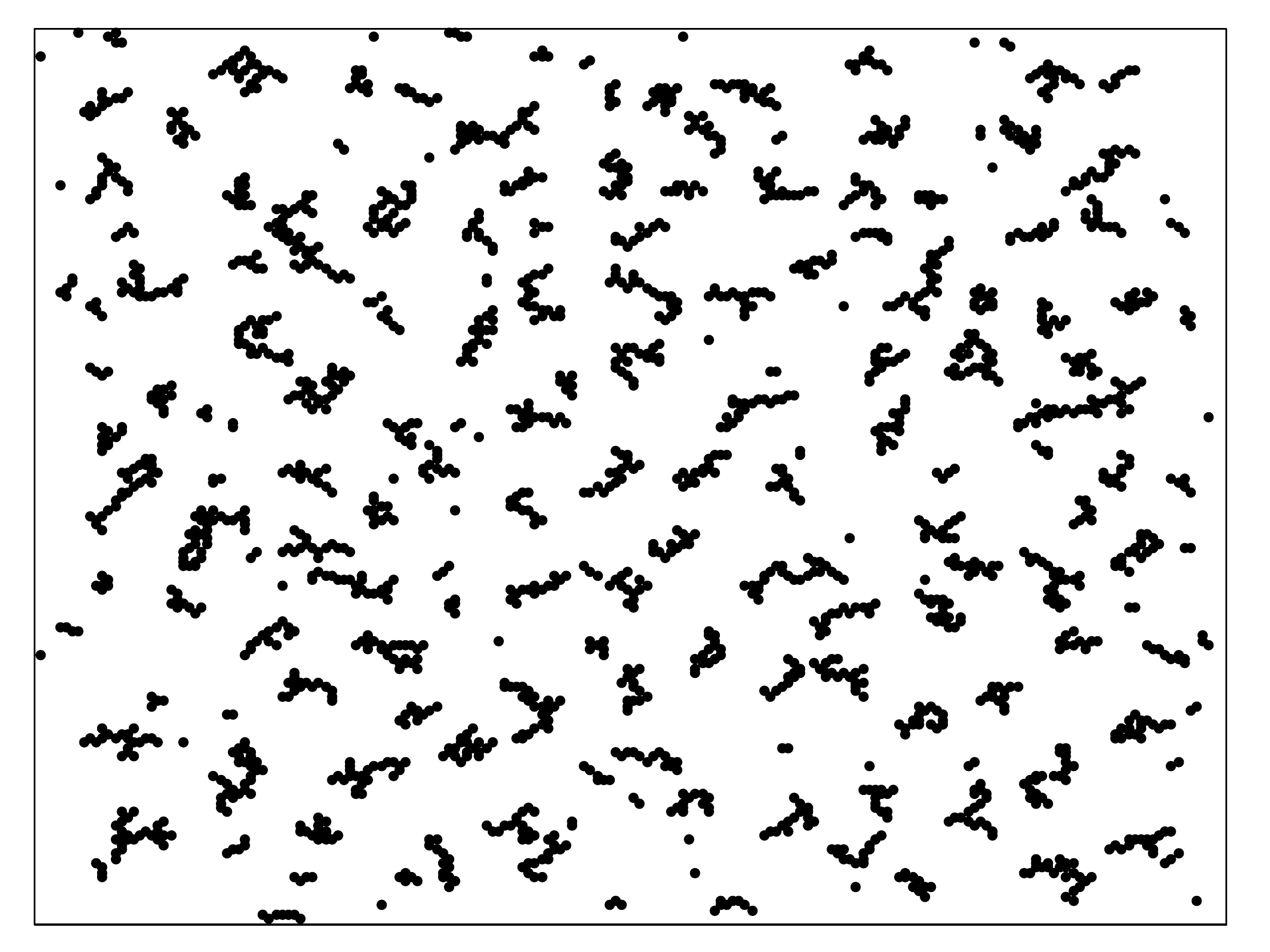}
\center {(b) T=100 }
\end{center}
\end{minipage}
\begin{minipage}[b]{0.3\textwidth}
\label{fig:cc}
\begin{center}
\includegraphics[scale=0.18]{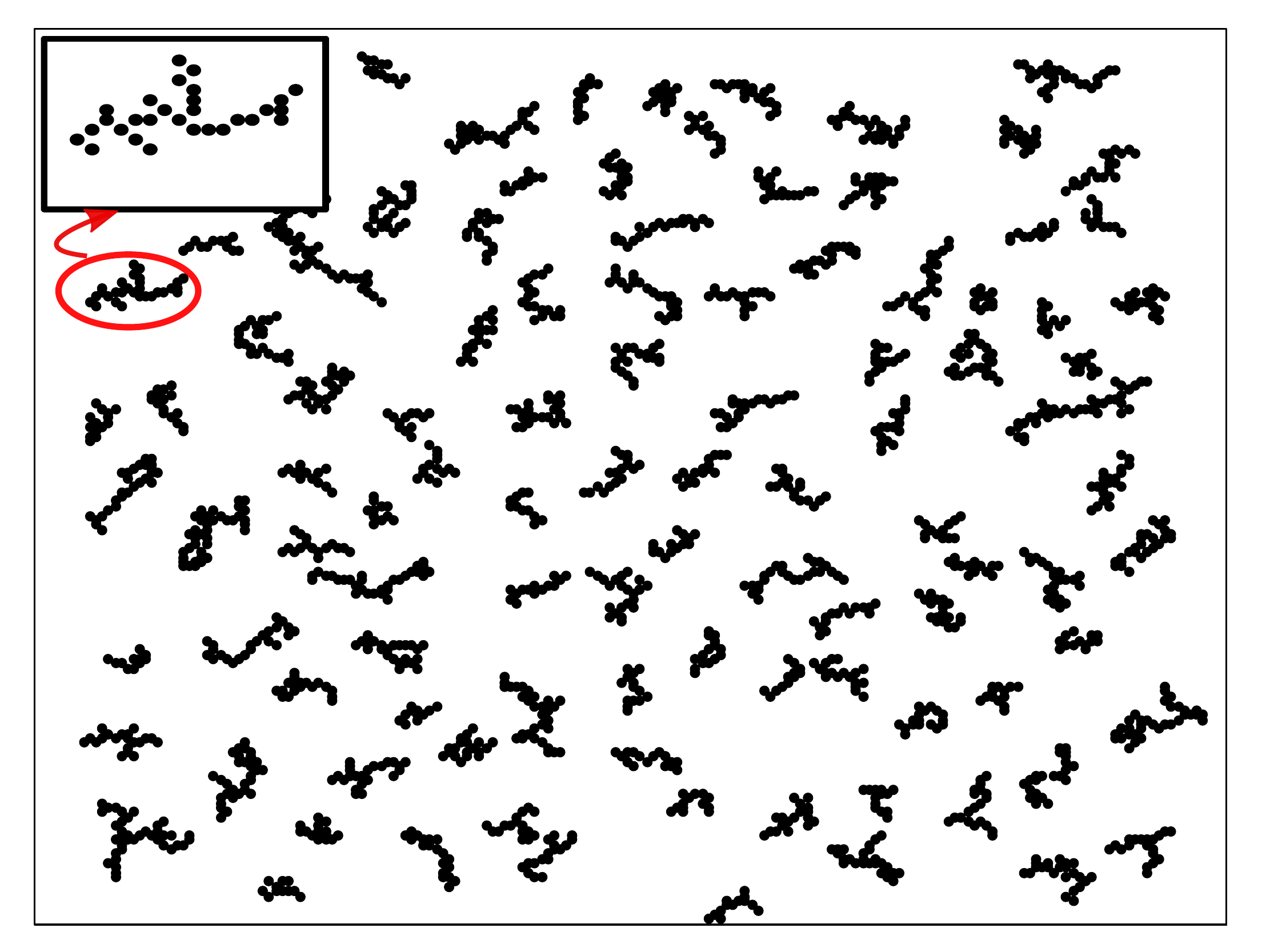}
\center {(c) T=5000}
\end{center}
\end{minipage}
\caption{{\bf Time evolution of diffusion limited aggregation of cells.} (a)  Presents a snapshot at simulation time $T=0$, here $2000$ cells are  deposited randomly on $(200 \times  200)$ lattice surface.
(b) While diffusing, cells collide and stick to each other to form aggregates, one such snapshot at $T=100$. (c) Shows irregular branched cellular aggregates at $T=5000$. The inset focuses one such irregular shaped cell aggregate.}
\label{fig:diff_20}
\end{figure}
\begin{figure}[!b]
\centering
\begin{minipage}[b]{0.3\textwidth}
\label{fig:cc}
%\caption{Using triangular lattice, \\A= 0.3994, D=1.601}
\begin{center}
\includegraphics[scale=0.18]{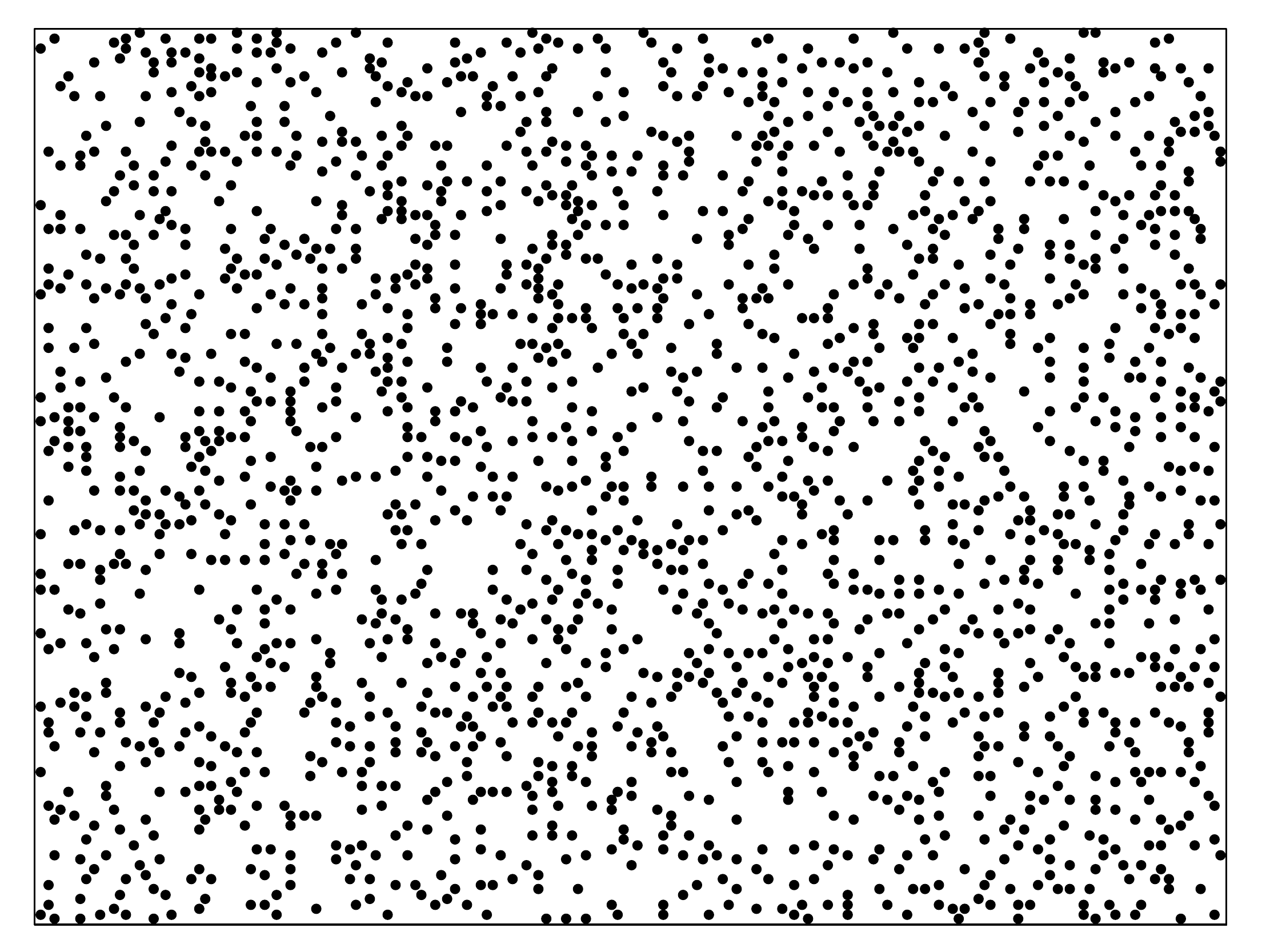}
\center {(a) T=0}
\end{center}
\end{minipage}
\begin{minipage}[b]{0.3\textwidth}
\label{fig:cc}
%\caption{Using square lattice, \\A= 0.39613, D=1.604}
\begin{center}
\includegraphics[scale=0.18]{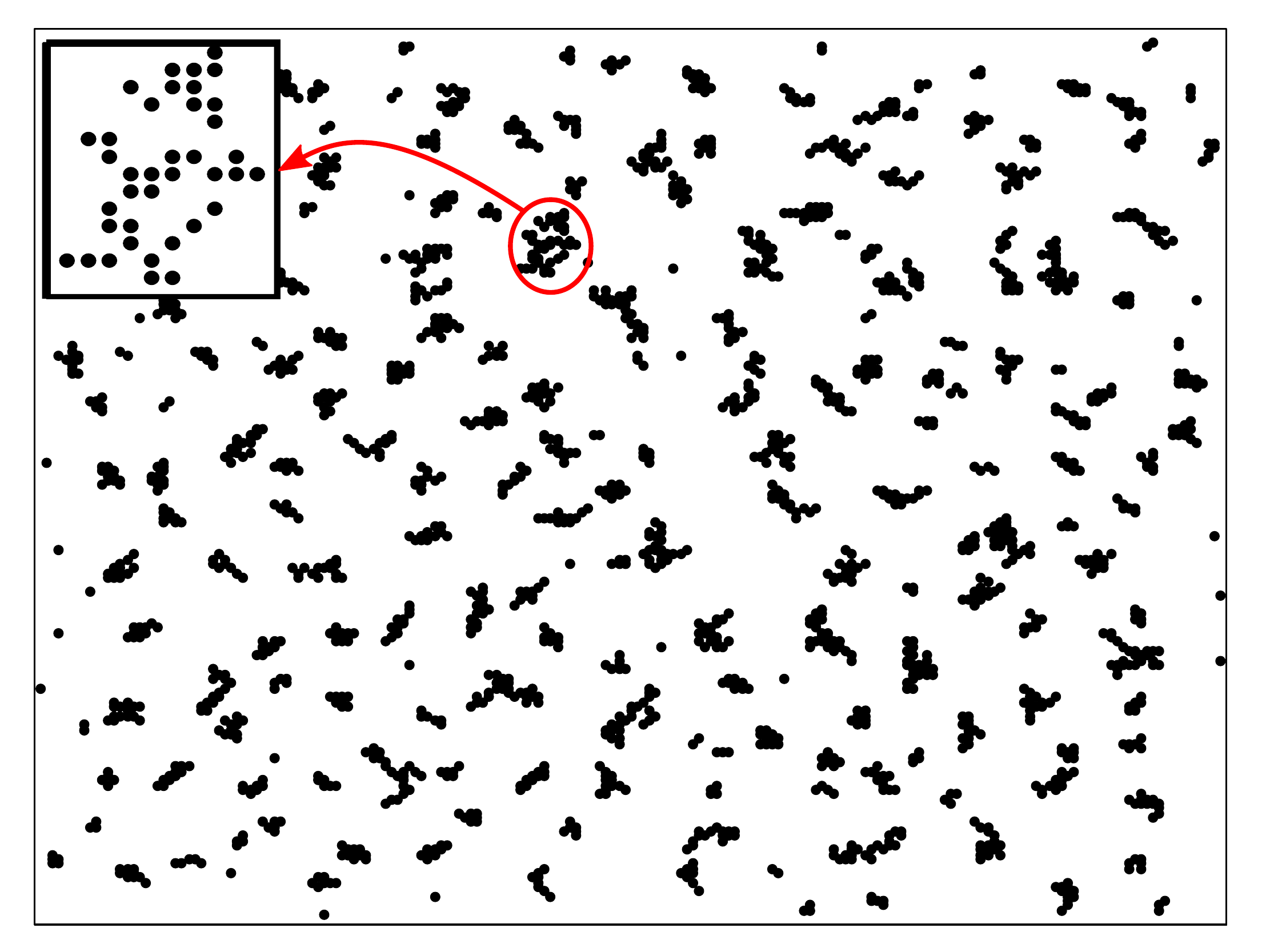}
\center {(b) T=100 }
\end{center}
\end{minipage}
\begin{minipage}[b]{0.3\textwidth}
\label{fig:cc}
\begin{center}
\includegraphics[scale=0.18]{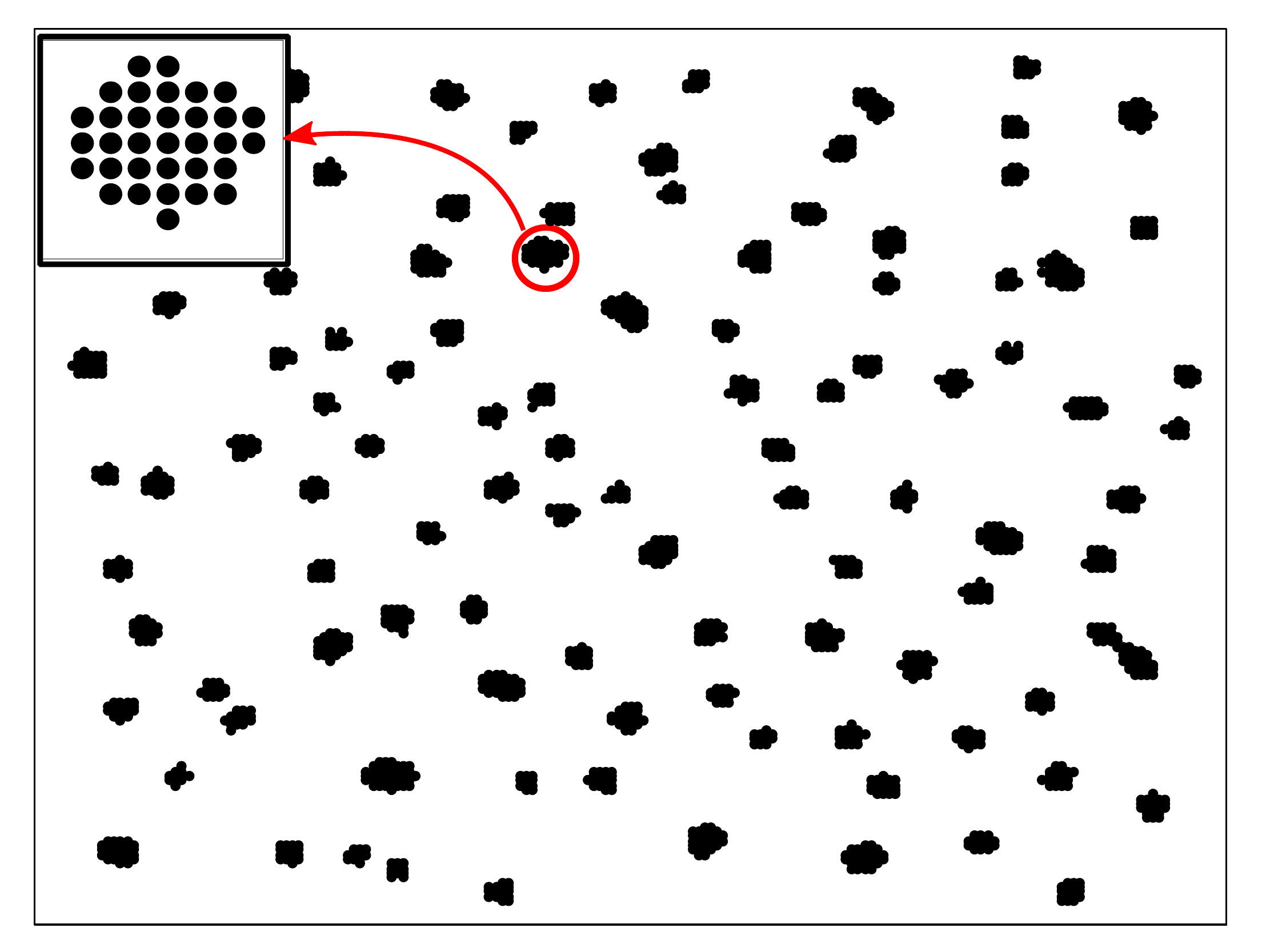}
\center {(c) T=4000}
\end{center}
\end{minipage}
\caption{{\bf Time evolution of aggregates formation due to diffusion and local reorganizations of cells.}
 (a) Presents a snapshot of random deposition of $2000$ cells on a $(200 \times 200)$ lattice  at simulation time $T=0$. (b) A snapshot of growing aggregates at $T=100$. Clusters initially grow as irregular shaped structure as shown in the inset. (c) Shows compact cellular aggregates after a sufficiently long time at $T=4000$. The compact shape develops due to rearrangement of clustering cells  to maximize the binding strength.  The inset shows such a compact cell aggregate. (Results presented here for $n_c=10$ and  $\beta=0.05$.)}
\label{fig:b05_r}
\end{figure}
\subsection{Aggregation of cells due to diffusion}

We, first, study the effect of simply diffusion into the aggregation of cells. 
The time evolution of cell aggregation due to diffusion is shown in 
Figs. \ref{fig:diff_20}(a)-(c). 
Figure \ref{fig:diff_20}(a) shows the initial deposition of $2000$ cells on a $200 \times 200$ lattice. These cells diffuse to any of its empty first nearest neighbouring sites with a probability $P_m(n)=\mu/n$. On the way, when they come across with other cells or clusters, they  stick irreversibly and form bigger clusters. Figure \ref{fig:diff_20}(b) shows such an intermediate time step of clustering of cells. As time progresses, these clusters grow with gathering of more diffusing clusters. However, the cluster motility gradually slow down with increase in cluster size and eventually stop moving after reaching a critical size, $n_c$, as observed in experiment \cite{Douezan2012}.
Figure \ref{fig:diff_20}(c) presents a snapshot of cellular aggregates acquiring irregular branched structures as predicted by diffusion limited aggregation studies \cite{Witten1983, Meakin1983, Meakin1984}. The irregular shapes arise, since the diffusing cells/clusters only access the protruding exterior regions as they irreversibly stick to the immediate neighbours, thus, do not roll down to the interior of the cluster.
Simulation results are presented here for $n_c=10$ as the critical size of the cluster is found  of the order of ten in experiment \cite{Douezan2012, Rodriguez2012}. We have also carried out simulations with $n_c=20$ and $30$; the size of aggregates becomes bigger and  develops more branched structure; however, the qualitative nature of aggregates remains the same.

\subsection{Aggregation: effect of local reorganisation of cells}
Next we investigate, in addition with diffusion, the influence of local reorganization of cells due to strong cell-cell binding affinity on the dynamics of aggregate formation.
As observed in experiments,  aggregates are grown generally compact in nature, 
where, cell-cell interaction to increase the cohesive binding strength plays a crucial role in smoothening out the irregular structures. 
We start with $N_0=2000$ cells placed randomly on  $2$D square lattice of size $L=200$.
In this case, cells or clusters  may diffuse with a probability, $P_m(n)=\mu/n$, otherwise may opt to locally reorganize with a rolling probability,  $P_r(n)=\beta n$ or just stay put as described in the previous section.
Figures \ref{fig:b05_r}(a)-(c) present one such simulation results of temporal evolution of aggregates formation (keeping $\mu=1$ and $\beta=0.05$). 
From these figures, it is clearly seen that as time progresses, aggregates grow in size  and  also gradually become compact due to local rearrangements of clustering cells. After a sufficiently long time, aggregates become fully compact and one such compact structure is shown in the inset of Fig. \ref{fig:b05_r}(c).

Moreover, we have carried out simulations for different cell density, 
$\rho=N_0/(L\times L)$, (keeping $N_0= 500$, $1000$, $2000$, $10000$, and $L=200$)  and  studied the effect of varying density in the formation of cellular aggregates.
Under low densities, cells/clusters diffuse for longer time to find another  cell or cluster to bind together and hence, the clustering process also takes longer time. However, for higher densities, diffusing  cells find other  cells  in its immediate vicinity. So, the growth process becomes faster and the cluster size  also gets bigger due to availability of more number of cells. On the other hand, having a large cluster size, local cell-cell rearrangements take longer time and thus, the overall compactification process becomes slower. 
Moreover, the compactification time strongly depends on the nature of cell types, cellular tendency of making and breaking adhesion bonds etc., in our simulations, which is represented by the variation in rolling coefficient, $\beta$. We discuss these  properties of the cellular aggregation in more details in the following subsections.

\subsection{Rate of cell aggregation}
In this section, we investigate the rate of aggregation, {\it i.e.}, number of clusters formation as a function of time. Clusters mainly form due to the random collisions of diffusing  clusters. Thus, at any instant, if there are $N(t)$ number of clusters  of single cell or multiple cells, then the occurrence of number of collisions for one cluster is $(N-1)$ as there are other $(N-1)$ surrounding clusters. Thus, at any time instant $t$, since all clusters randomly move and collide with other; so the total number of collisions considering all clusters movement is proportional to $N(N-1)$. Thus, the time evolution of the number of clusters can be described by the equation,

%\begin{subequations}
\begin{equation}
\frac{d N(t)}{dt}=-KN(t)(N(t)-1)\approx -KN(t)^2, 
\label{eq_cluster_evolution}
\end{equation}
where $K$ is the rate constant which depends on the size and motility of the cell  type. \cite{Douezan2012}

\begin{figure}[!h]
\centering
%\caption{Using square lattice, \\A= 0.39613, D=1.604}
\begin{minipage}[b]{0.42\textwidth}
\begin{center}
\includegraphics[scale=0.65]{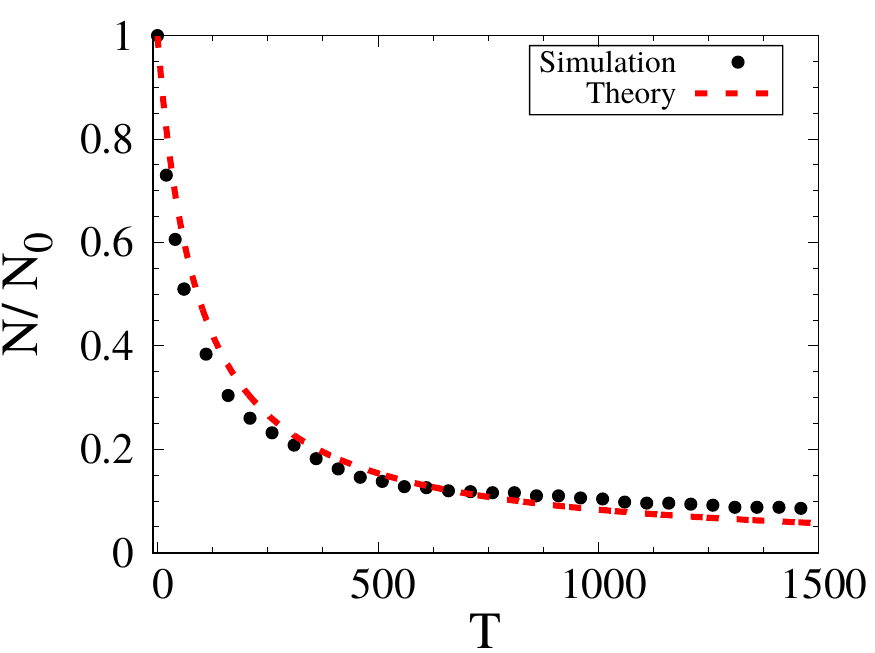}
\center {(a)}
\end{center}
\end{minipage}
\begin{minipage}[b]{0.42\textwidth}
%\caption{Using triangular lattice, \\A= 0.3994, D=1.601}
\begin{center}
\includegraphics[scale=0.65]{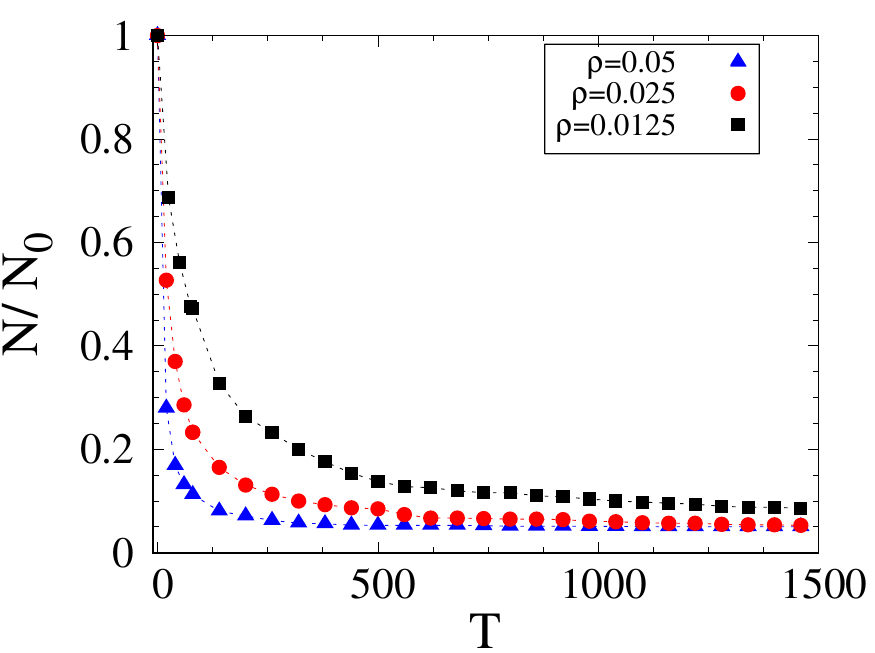}
\center {(b)}
\end{center}
\end{minipage}
\caption{{\bf Time evolution of number of clusters.}(a) Comparison of simulation result of rate of aggregation ($(\protect\markernine)$, keeping $\rho=0.0125,  \beta=0.05$) with theoretical model prediction, solution of Eq. \ref{eq_cluster_evolution}, shown by the dashed curve.   (b) Formation of number of aggregates as a function of time for different  initial cluster density, $\rho=0.0125 (\protect\markereight)$, $0.025 (\protect\markerfive)$, and $0.05 (\protect\markerseven)$.
}
\label{fig:peri_c_diam}
\end{figure}

From our simulation, we evaluate the rate of cell aggregation, {\it i.e.}, the rate of change of number of clusters during the aggregation process.  We find as the  clusters while diffusing collide with each other and merge to form bigger aggregates and hence, the number of clusters decreases with time.
Besides, since the aggregate stops moving after reaching a critical size (as its mass gets heavier), number of clusters eventually reaches to a steady value. 
Figure \ref{fig:peri_c_diam}(a) shows our simulation results of time evolution of number of clusters which is in good agreement with recent experimental observations \cite{Douezan2012} (as it is also seen from Figure \zref{fig:cluster_num} provided in the supporting information).
We also compare our simulation results with the theoretical prediction of Eq. (\ref{eq_cluster_evolution}). As seen  from Fig. \ref{fig:peri_c_diam}(a), the  theoretical prediction agrees quite well with our simulations as well as with the experimental observations by tuning only one fitting parameter, the rate constant $K$.

Moreover, we also study the effect of cell density variation on the rate of cell aggregation. Figure \ref{fig:peri_c_diam}(b) shows the simulation results for three different  densities $\rho=0.0125$, $0.025$, and $0.05$ (with initial cell  number, $N_0=500$, $1000$, $2000$ and $L=200$). As shown in Fig. \ref{fig:peri_c_diam}(b), for higher density, the decrease in the number of clusters is faster; since  chances of finding neighbouring cells/clusters are higher, the rate of merging of clusters is also high. In other words, the growth of aggregates happens faster for higher cell density.

\subsection{Evolution of surface area of an aggregate}
 Resent  experimental studies show that the surface area of the growing aggregates on non-adhesive substrate exhibits a non-monotonic evolution  unlike the monotonic increase in the area as observed in case of spreading of cell aggregates \cite{Douezan2011}. 
In order to get an insight into the dynamics, we also investigate the area evolution of the aggregates. In our simulations, we  consider two methods to estimate the surface area of an aggregate. In one method, we calculate the perimeter of the cluster, which in turn provides an estimation of the projected area and in another, we calculate the radius of gyration of the cluster to evaluate  the surface area (details have been given as supporting materils).

\begin{figure}[!t]
\begin{minipage}[b]{0.5\textwidth}
%\caption{Using square lattice, \\A= 0.39613, D=1.604}
\begin{center}
\includegraphics[scale=0.75]{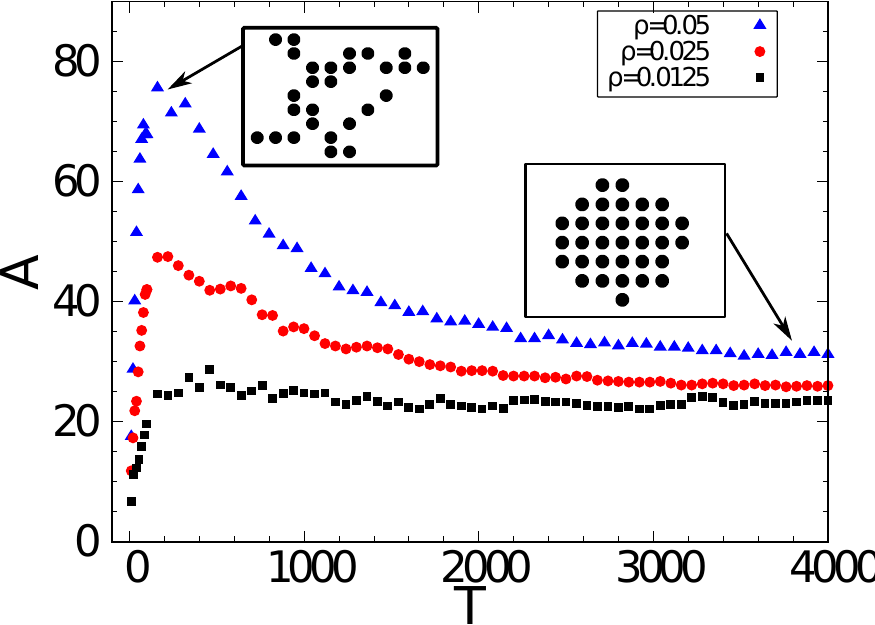}
\center {(a)}
\end{center}
\end{minipage}
\begin{minipage}[b]{0.5\textwidth}
%\caption{Using triangular lattice, \\A= 0.3994, D=1.601}
\begin{center}
\includegraphics[scale=0.75]{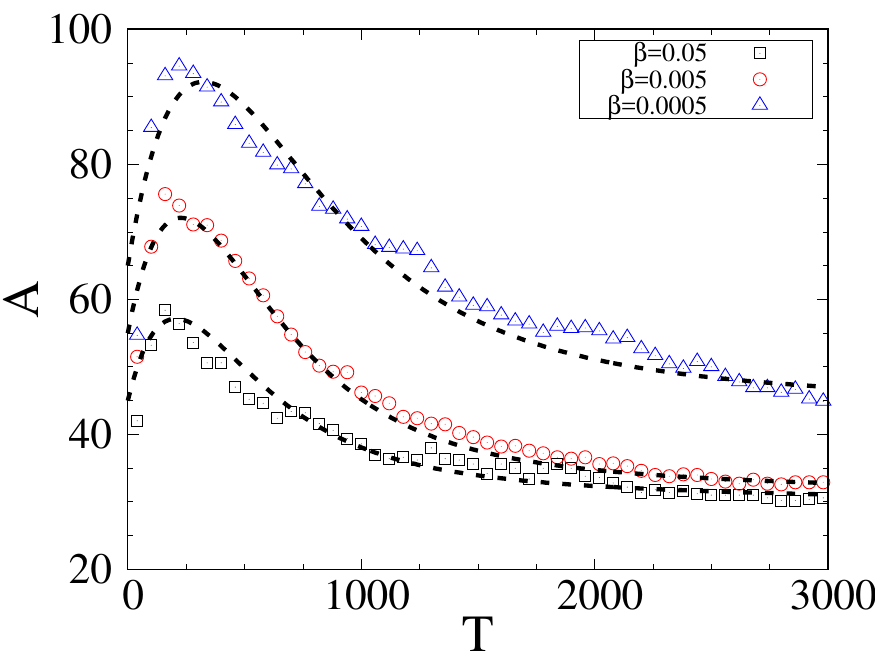}
\center {(b)}
\end{center}
\end{minipage}
\caption{ {\bf Time evolution of the surface area of cellular aggregate} (a) Simulation results for different initial cluster density, $\rho=0.0125 (\protect\markereight)$, $0.025(\protect\markerfive)$, $0.05(\protect\markerseven)$. Left panel inset shows the development of irregular shaped structure at some initial time and right panel inset shows the compactified structure at a longer period of time. (b) Shows the effect of rolling coefficient on the surface area evolution for $\beta=0.0005 (\protect\markerten)$, $0.005 (\protect\markertwelve)$, and $0.05 (\protect\markereleven)$.
Theoretical prediction, solution of Eq. (\ref{eq:theory_area}), is shown by the dashed curve ($\tau_c$ has been varied for different $\beta$ values and while keeping $K$ constant).
}
\label{fig:peri_c_diag}
\end{figure}

We have studied the time evolution of the surface area averaged over many aggregates for different cell density, $\rho=0.0125, 0.025, 0.05$  and rolling tendency of varied cell type given by $\beta = 0.0005, 0.005, 0.05$ as shown in Figure \ref{fig:peri_c_diag} (a) and \ref{fig:peri_c_diag}(b) respectively. 
 We find, as observed in experiment, \cite{Douezan2012} at the early stages of the growth process, the surface area of the cluster increases rapidly due to collisions and merging of clusters. However, once the aggregate reaches a critical size, it stops moving and other than occasional joining of randomly wandering of small clusters, the cells reorganize among themselves to be surrounded by the maximum possible neighbours, and hence, the surface area start decreasing due to the compactification process and finally reaches to a steady state value giving rise to the most compact structure.   
Insets of Fig. \ref{fig:peri_c_diag} (a) show an irregular shaped structure that arises due to joining of diffusing cells at early times and an compact structure due to  self-organization of the clustering cells at a longer period of time.
Importantly, our model could capture the non-linear nature of the cluster area evolution as found in experiments which could also be seen from Figure \zref{fig:area} in the supplementary information.

The time evolution of the cluster area, $A(t)$,  during the aggregation process can also be theoretically modeled following earlier studies \cite{Koch1990, Hiram1983} as,
\begin{equation}
\frac{dA(t)}{dt}=KN(t)A(t)-\frac{1}{\tau_c}[A(t)-A_{final}];
\label{eq:theory_area}
\end{equation} 
where $N(t)$ is the number of clusters at any instant $t$. Here, 
the first term represents the area increase due to diffusive collision and merging of clusters and the second term represents the decrease in area with a characteristic relaxation time, $\tau_c$, which is related to the tendency of reorganization of cells to reach to the most compact structure of an area $A_{final}$.

 We now compare simulation results of our cellular automata model with the theoretical prediction.
%Eq. (\ref{eq:theory_area}). 
%as shown in Fig. \ref{fig:peri_c_diag} (b).
We numerically solve Eq. (\ref{eq:theory_area}) substituting $N(t)$ from Eq. (\ref{eq_cluster_evolution})  keeping $K$ constant and varying values of $\tau_c$  and then compare it with different values of $\beta$ ($=0.05$, $0.005$, and $0.0005$). 
%\textbf{Here $\beta$ actually controls the time scale of compaction process.} 
As seen from the Fig. \ref{fig:peri_c_diag} (b), in our simulation, the rolling coefficient, $\beta$, plays the role as of $1/\tau_c$ in Eq. (\ref{eq:theory_area}). For low  values of $\beta$, since the chances of cellular rolling/reorganization decreases,  sparse branching structure prevails for longer period of time,
thus, the coalescence time, $\tau_c$  becomes higher. 
On the other hand, increasing $\beta$ value increases the rate of coalescence and thus, the cluster compactifies faster.

\subsection{Compactness of cellular aggregate}
We further investigate how the cellular cohesive strength  evolves as the cluster grows in time. 
In our simulations, it is measured by estimating the adhesion junctions formed between cells, {\it i.e.}, the total number of bonds formed among the constitutive cells within the aggregate. 
As time progresses, the binding strength increases due to more number of cells/clusters joining the aggregate. On the other hand, addition of new cells in the cluster due to diffusion makes the aggregate to spread out. However, aggregates slowly become compact due to local reorganization of cells within the cluster to
minimize the binding free energy by forming maximum possible bonds with the neighbouring cells. In our model, we define the compactness of an aggregate, $C_{r}$, as total number of bonds in a cluster.  
It has been, further, normalized to differentiate the increase in number of bonds due to addition of diffusing cells or due to local rearrangement of the relative position of the cells to  maximize the binding strength as defined by $C_{nr}$,
\begin{equation}
C_{nr}=\frac{Total\ no.  \ of  \ bonds  \ in   \ a  \ cluster \ of \ n \ cells}{Bonds \ for \ making \ 1D \ chain \ of \ n \ cells}
\end{equation}
Thus,  $C_{r}$ has information  about the total number of bonds in the aggregate, on the other hand, $C_{nr}$ represents that given a number of cells in an aggregate, how closely cells are packed together.

\begin{figure}[!h]
\begin{minipage}[b]{0.5\textwidth}
%\caption{Using square lattice, \\A= 0.39613, D=1.604}
\begin{center}
\includegraphics[scale=0.75]{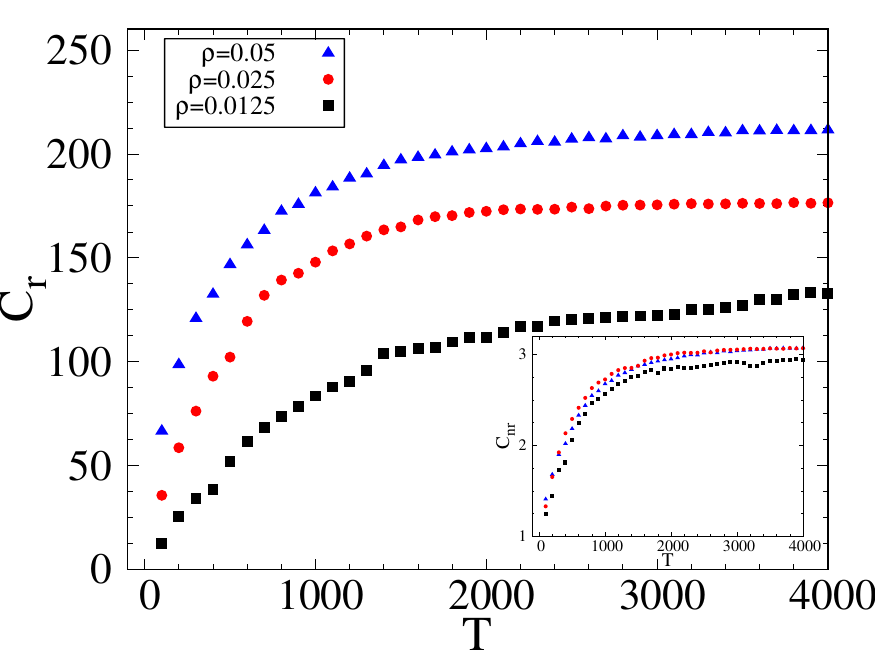}
\center {(a)}
\end{center}
\end{minipage}
\begin{minipage}[b]{0.5\textwidth}
%\caption{Using triangular lattice, \\A= 0.3994, D=1.601}
\begin{center}
\includegraphics[scale=0.75]{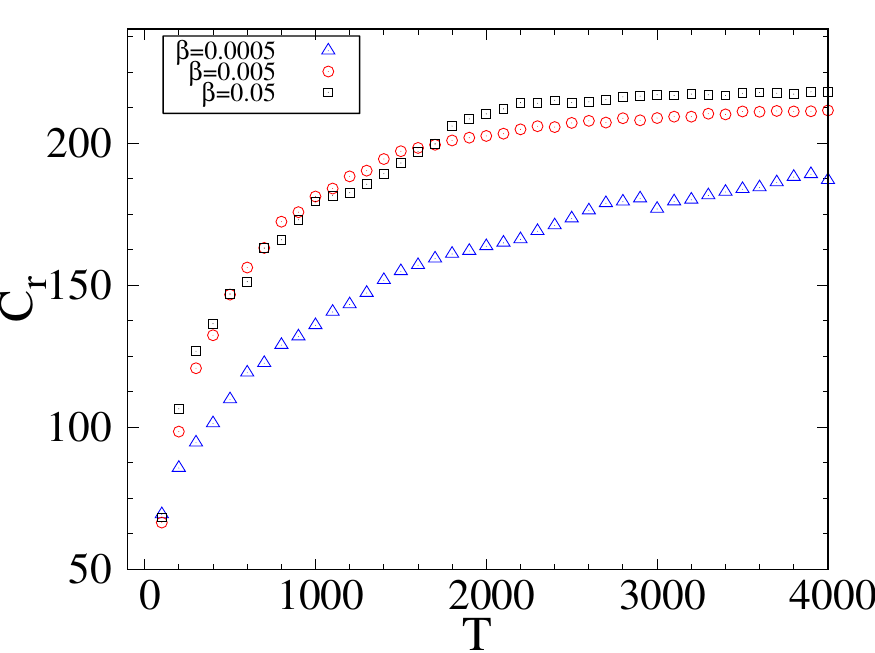}
\center {(b)}
\end{center}
\end{minipage}
\caption{ {\bf Time evolution of the compactness of cellular aggregate.} (a) Shows the effect of different  initial cluster density, $\rho=0.0125 (\protect\markereight)$, $0.025 (\protect\markerfive)$, and $0.05 (\protect\markerseven)$. Corresponding $C_{nr}$, for different density is shown in inset.(b) Plot shows the dependence of the aggregate compactness on the variation of rolling coefficient for $\beta=0.0005 (\protect\markerten)$, $0.005 (\protect\markertwelve)$, and $0.05 (\protect\markereleven)$.}
\label{fig:compt_c_diag}
\end{figure}

We have simulated the time evolution of the compactness, $C_r$, of an aggregate for different cell  density, $\rho$ and varying rolling coefficient, $\beta$ averaged over many aggregates  shown in Figs. \ref{fig:compt_c_diag}(a) and (b) respectively. 
As seen from the plot, the compactness, {\it i.e.}, the binding strength of  an aggregate increases as the cluster grows with time. The initial rapid increase is due to joining of diffusing clusters then the increase happens mainly due to the local reorganization of cells that involves breaking and making of new bonds; thus, it occurs at a much slower rate compare to diffusion.
Moreover, as cell density increases, the number of cells in an aggregate also increases and hence, the total number of bonds, {\it i.e.}, compactness also becomes higher. 
Further, as seen from the inset of Fig. \ref{fig:compt_c_diag}(a), since $C_{nr}$ is normalized by the cluster size, all curves for different cell density collapse into a single curve.

\subsection{Fractal geometry of the aggregate}
\begin{figure}[!b]
\begin{center}
\includegraphics[scale=0.8]{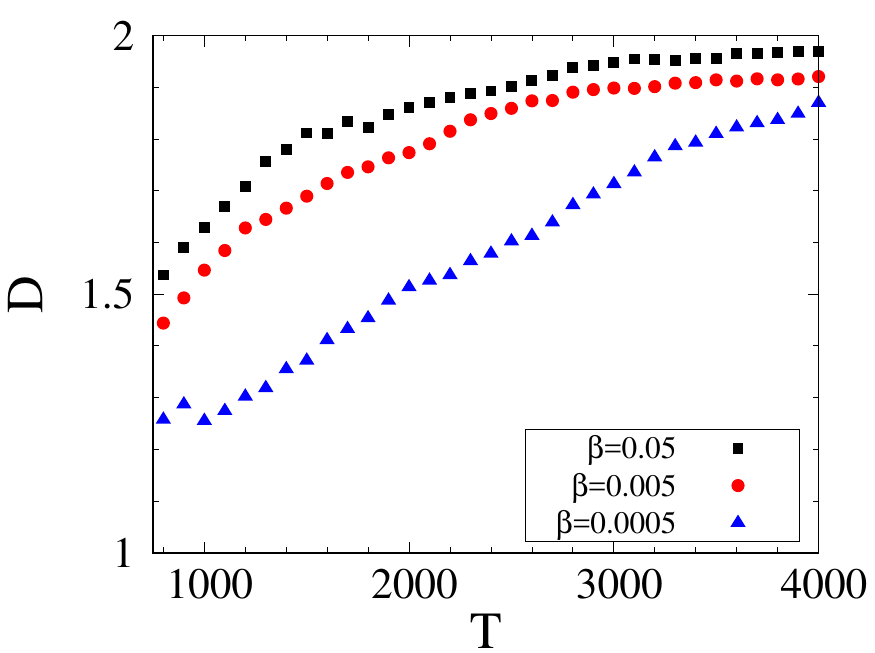}
\center
\end{center}
\caption{Time variation of fractal dimension of a growing cellular aggregate for 
different rolling coefficient, $\beta=0.0005 (\protect\markerseven)$, $0.005 (\protect\markerfive)$, and $0.05 (\protect\markereight)$. Here, the initial cluster density is kept constant at $\rho=0.05$.
} 
\label{fig:fractal}
\end{figure}

Recent studies have shown that the fractal geometry can be useful to understand the underlying mechanisms of tissue growth  as living tissues are spatially heterogeneous and thus, exhibit fractal pattens \cite{Baish2000}. Moreover, it has been observed that  fractal analysis may provide an efficient way to distinguish normal versus cancerous tissue growth.
We have, therefore, characterized the fractal dimension of growing aggregates to get insights into the complexity of the structure.
In our simulations, the fractal dimension, $D$, has been estimated from the radius of gyration, $R_g$, of the evolving cluster using the following relationship \cite{Meakin1983, Kolb1983},
\begin{equation}
%\begin{align}
R_g \sim n^{1/D};
%R_g = C n^{1/D}
%\end{align}
\end{equation}
where $n$ is the number of cells in the cluster.
As discussed in the previous section, at early times, the cluster grows due to numerous collisions between the diffusing cells/clusters; thus initially, it develops a branched sparse structure and gradually self-organization of clustering cells  gives rise to the compact structure. In our simulations, we start calculating the fractal dimension  after the aggregate has grown to a branched structure and then study as time progresses, how the dimension changes due to the compactification process. The time evolution of the fractal dimension, $D$, averaged over many such aggregates  is presented in Fig. \ref{fig:fractal}.
As seen from the figure, at the initial stage, the fractal dimension turns out to be similar to the diffusion limited aggregates \cite{Witten1983, Meakin1983, Sander1994}, however, it slowly increases as the clustering cells relocate to the energetically favourable binding sites, and as a result, also compactifies the aggregate. Eventually, the fractal dimension approaches to two for the most compact aggregated structure, as it is expected since the simulation is done on a $2$D surface. 
Moreover, as seen from Fig. \ref{fig:fractal}, with increase in rolling coefficient, $\beta$, cluster cells to have higher chances to find their favourable binding sites; thus the compaction process becomes quicker and hence, the fractal dimension of the aggregate also increases faster. 
 On the other hand, it has been it has been observed that
the fractal dimension of cancerous tissue increases with increase with progress in cancer stages as the heterogeneity changes due to accumulation of more masses. Interestingly, our study reveals that 
the difference in the tendency of cell types to  increase binding strength also plays a crucial role in determining the structure of the tissue pattern which can be tested further by suitable experiments.

\section{Conclusion}
 We have developed a cellular automata model to study the aggregation dynamics of a seemingly disordered tissue cell population in the absence of any external mediator. 
This model  based on the active motility and local reorganization of cells could  successfully capture the structural and temporal evolution of aggregates that leads to the compact tissue structures as observed in experiments.
Importantly, our study shows that the sole consideration of the cellular tendency of forming bonds  with the neighbouring cells to maximize cohesive strength is sufficient for the spontaneous emergence of compact tissues.
Moreover, it provides several insights into the dynamics of the cell aggregation process. 
It reveals the existence of two distinct time scales - one fast time scale associated with the diffusion of cells and  another much slower time scale associated with the tissue compaction process that involves breaking of cell-cell cohesive bonds and making of  new bonds
leading to local reorganization of cells.
Besides, as found in experiments, our simulation results also successfully predicts many dynamical properties of the growing aggregates, such as, the rate of cell aggregation, the non-linear  evolution of the surface area, the binding strength and the compactness of the growing aggregate \cite{Douezan2012,  Rodriguez2012, Guo2006}. 
 Moreover, we have demonstrated that the variation in tendency of rolling and reorganization of cells have a profound effect on the formation of tissue shapes and structures and it could be further tested by carrying out suitable experiments. Our theoretical model in essence is of a generic nature and hence 
can  be extended to other systems with suitable modifications. 
Moreover, since tissue development is quite a complex process and
current tissue engineering procedures are still very experimental and also expensive; thus, simple theoretical and computational model studies are envisaged to facilitate the understanding of how individual cells organize into tissues much like as it has been done in passive growth processes which once seemed to be a difficult prospect.

\section{Acknowledgments}
The authors acknowledge the financial support from Science and Engineering Research Board (SERB), Grant No. SR/FTP/PS-105/2013, Department of Science and Technology (DST), India.

%\nolinenumbers

% Either type in your references using
% \begin{thebibliography}{}
% Text
% \end{thebibliography}
%
% or
%
% Compile your BiBTeX database using our plos2015.bst
% style file and paste the contents of your .bbl file
% here. See http://journals.plos.org/plosone/s/latex for 

%\bibliography{cell_agg_draft_pre2}

%\begin{thebibliography}{10}

%\providecommand{\noopsort}[1]{}\providecommand{\singleletter}[1]{#1}%

\end{document}